\documentclass[twocolumn]{IEEEtran}

\usepackage{amsmath,bm}
\usepackage{amsthm}
\usepackage{amssymb}
\usepackage{graphicx}
\usepackage{epsfig}
\usepackage{psfrag}
\usepackage{subfigure}
\usepackage{url}
\usepackage{stfloats}
\usepackage{amsmath}
\usepackage{algorithm}
\usepackage{algorithmic}
\usepackage{diagbox}
\usepackage{dsfont}
\usepackage{bbding}

\usepackage{amsmath}
\usepackage{graphics}
\usepackage{graphicx}
\usepackage{amssymb}
\usepackage{algorithm}
\usepackage{algorithmic}
\usepackage{mathrsfs}
\usepackage{booktabs}
\usepackage{bm}
\usepackage{bbm}
\usepackage{float}
\usepackage{url}
\usepackage{amsfonts}
\usepackage[square,sort&compress,numbers]{natbib}
\usepackage{multirow}
\usepackage{textcomp}
\usepackage{supertabular}
\usepackage{array}
\usepackage{color,soul}

\begin{document}

\title{Multi-grained Attention Networks for Single Image Super-Resolution}

\author{ {Huapeng Wu, Zhengxia Zou, Jie Gui, \textit{Senior Member, IEEE}, Wen-Jun Zeng, Jieping Ye, \textit{Senior Member, IEEE}, Jun Zhang, \textit{Member IEEE}, Hongyi Liu, \textit{Member IEEE}, and Zhihui Wei}

\IEEEcompsocitemizethanks{
\IEEEcompsocthanksitem This work was supported in part by the National Natural
Science Foundation of China under Grant 11431015, Grant 61671243, Grant 61971223 and Grant 61572463. Corresponding author: Zhihui Wei (e-mail: gswei@njust.edu.cn).
\IEEEcompsocthanksitem Huapeng Wu is with the School of Computer Science and Engineering, Nanjing University of Science and Technology, 210094, China. 
\IEEEcompsocthanksitem Zhengxia Zou, Jie Gui and Wen-Jun Zeng are with the Department of Computational Medicine and Bioinformatics, University of Michigan, Ann Arbor, MI, 48109 U.S.A.
\IEEEcompsocthanksitem Jieping Ye is with the Department of Computational Medicine and Bioinformatics, and the Department of Electrical Engineering and Computer Science, University of Michigan, Ann Arbor, MI 48109, U.S.A., and with the DiDi AI Labs, DiDi Chuxing, Beijing, 100085, China.
\IEEEcompsocthanksitem Jun Zhang and Hongyi Liu are with the School of Science, Nanjing University of Science and Technology, 210094, China.
\IEEEcompsocthanksitem Zhihui Wei (corresponding author: gswei@njust.edu.cn) is with the School of Computer Science and Engineering, Nanjing University of Science and Technology, 210094, China, and with the School of Science, Nanjing University of Science and Technology, 210094, China.

} 
} 

\markboth{}%
{Shell \MakeLowercase{\textit{et al.}}: Bare Demo of IEEEtran.cls for Journals}

\maketitle

\newcolumntype{L}[1]{>{\raggedright\arraybackslash}p{#1}}
\newcolumntype{C}[1]{>{\centering\arraybackslash}p{#1}}
\newcolumntype{R}[1]{>{\raggedleft\arraybackslash}p{#1}}

\begin{abstract}
Deep Convolutional Neural Networks (CNN) have drawn great attention in image super-resolution (SR). Recently, visual attention mechanism, which exploits both of the feature importance and contextual cues, has been introduced to image SR and proves to be effective to improve CNN-based SR performance. In this paper, we make a thorough investigation on the attention mechanisms in a SR model and shed light on how simple and effective improvements on these ideas improve the state-of-the-arts. We further propose a unified approach called ``multi-grained attention networks (MGAN)'' which fully exploits the advantages of multi-scale and attention mechanisms in SR tasks. In our method, the importance of each neuron is computed according to its surrounding regions in a multi-grained fashion and then is used to adaptively re-scale the feature responses. More importantly, the ``channel attention'' and ``spatial attention'' strategies in previous methods can be essentially considered as two special cases of our method. We also introduce multi-scale dense connections to extract the image features at multiple scales and capture the features of different layers through dense skip connections. Ablation studies on benchmark datasets demonstrate the effectiveness of our method. In comparison with other state-of-the-art SR methods, our method shows the superiority in terms of both accuracy and model size.

\end{abstract}

\begin{IEEEkeywords}
Super-resolution, convolutional neural networks, multi-grained attention, multi-scale dense connections.
\end{IEEEkeywords}

\section{Introduction}

Image Super-Resolution (SR) is an important image processing technique that recovers high-resolution images from low-resolution ones. Image SR has drawn great attention recently in a variety of research fields, such as remote sensing imaging \cite{wang2014compressed}, medical imaging \cite{shi2013cardiac}, and video surveillance \cite{jiang2016srlsp}. Image SR is a representative of the ill-posed inverse problems. This is because information is lost during the degradation process and each Low-Resolution (LR) image may correspond to multiple High-Resolution (HR) ones. Great efforts have been expended for this problem, where the previous methods can be divided into three groups: the interpolation-based methods \cite{zhou2012interpolation}, the reconstruction-based methods \cite{zhang2012single}, and the learning-based methods \cite{wu2018high, dong2015image, kim2016accurate, kim2016deeply, tai2017image, lai2017deep, ledig2017photo, lim2017enhanced}.

In recent years, the fast development of deep learning technology \cite{lecun2015deep} has greatly advanced the research of single image SR. Due to its high-level representation and strong learning ability, the deep convolutional neural networks (CNN) have soon become the \textit{de facto} framework for the SR community. The first attempt to use CNN for single image SR was made by Dong \emph{et al}.\ \cite{dong2015image}. They built a three-layer CNN model called ``SRCNN'' to learn a nonlinear mapping from LR to HR image pairs in an end-to-end fashion, which shows significant improvement over most conventional methods. Later, some other researchers proposed much deeper networks by integrating the idea of residual learning \cite{kim2016accurate, kim2016deeply, tai2017image} and recursive learning \cite{kim2016deeply, tai2017image}, achieving substantial improvement over the SRCNN. However, these methods firstly upscale a LR input images to the desired output size by using bicubic interpolation before it is fed into the networks, thus leading to extra computational cost and reconstruction artifacts. To overcome this problem, some new methods are proposed to upsample spatial resolution in the latter layers of the network. Dong \emph{et al}. \cite{dong2016accelerating} proposed a deconvolution operation for upscaling the final LR feature maps. Shi \emph{et al}. \cite{shi2016real} introduced a more effective sub-pixel convolution layer to upscale the LR feature maps to the size of HR images at the output end of the network. Then, this efficient post-processing method is widely used in SR \cite{lim2017enhanced, zhang2018image}, which not only further increases the depth of the network, but also reduces computational load. In a recent work proposed by Lee \emph{et al}  \cite{lim2017enhanced}, the SR model is built with over 160 layers. They employed a simplified ResNet \cite{he2016deep} architecture by removing the normalization layers in the SRResNet \cite{ledig2017photo} and won the championship of the NTIRE2017 super-resolution challenge \cite{timofte2017ntire}. 

\begin{figure*}
  \centering
  \includegraphics[width=\linewidth]{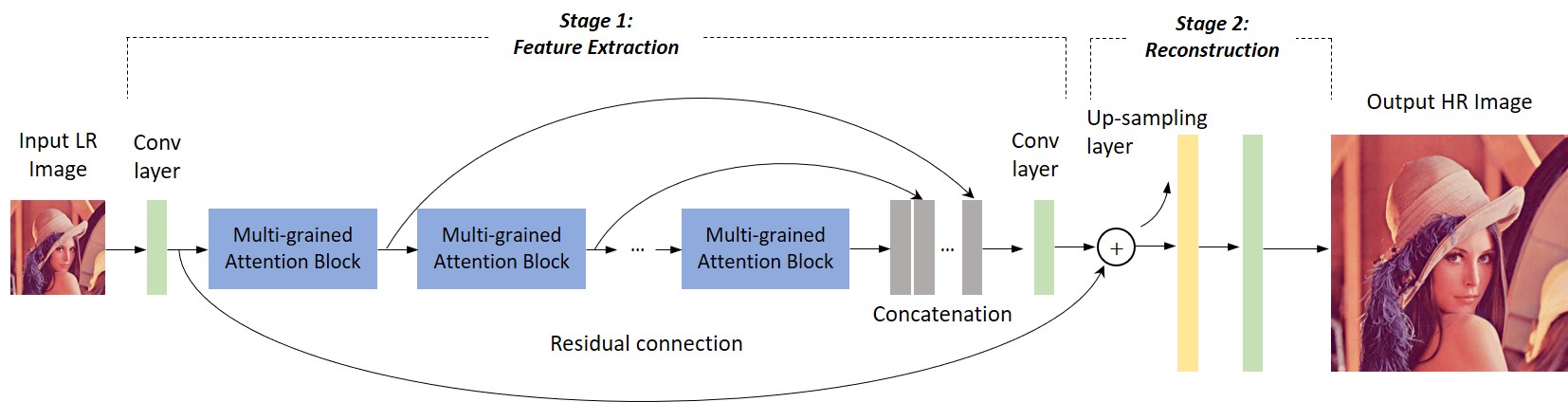}\\
  \caption{An overview of our multi-grained attention network (MGAN). Our network consists of several multi-grand attention blocks. To further exploit the information of hierarchical features \cite{li2018multi, zhang2018residual}, the features produced by different blocks are concatenated together to obtain the final representation.}\label{fig:architecture}
\end{figure*}

More recently, the idea of multi-scale feature fusion has been introduced to SR, where most of them are based on inception mechanism in GoogLeNet \cite{szegedy2015going}. 
In image SR, it has been demonstrated that making full use of multi-scale features improves the restoration results \cite{zhang2012multi}.  Specially, Li \emph{et al}.\ \cite{li2018multi} employed inception architecture to construct multi-scale residual block as the building block for the SR model. However, like many deep learning networks such as VDSR \cite{kim2016accurate}, LapSRN \cite{lai2017deep}, EDSR \cite{lim2017enhanced}, this method \cite{li2018multi} ignores the integration of the information from different convolutional layers for image reconstruction. Although the above methods achieve incremental results by utilizing the dense skip connections \cite{huang2017densely} to enrich their features, e.g., SRDenseNet \cite{tong2017image}, RDN \cite{zhang2018residual} and CSAR \cite{hu2019channel}, they ignore the importance of multi-scale features. Besides, some other methods like MemoryNet \cite{tai2017memnet} use a similar approach by combining information from the preceding memory blocks. However, it takes the pre-sampled images to form input which results in additional computational overhead. 

\textbf{Attention mechanism in image SR.} More recently, the ``attention mechanism'' has been introduced to single image SR and proves to be effective to improve the performance of a deep CNN based SR model. The attention mechanism was originally proposed in machine translation to improve the performance of a RNN model by taking into account the input from several time steps to make one-step prediction \cite{bahdanau2014neural}. In a CNN-based model, the introduction of an attention module is helpful for investigating the correlations of different feature channels and spatial locations, which has now been widely used in many computer vision tasks, such as object detection \cite{zhang2018occluded}, optical character recognition \cite{wojna2017attention}, and image captioning \cite{xu2015show}. Although CNNs have shown great superiority in SR tasks, its drawbacks are obvious. As the convolution operation in a standard CNN model is translation invariant, a convolution layer treats different feature locations equally and thus is hard to learn effective representations to exploit contextual cues. 

To this end, some researchers introduce the attention mechanism into a SR model to improve the features \cite{zhang2018image, cheng2018sesr, hu2019channel}. Zhang \emph{et al}.\ \cite{zhang2018image} introduced the Squeeze and Excitation (SE) module \cite{hu2018squeeze} (a.k.a the channel attention module) into a SR model to re-weight the importance of each feature channel by learning the interdependencies among channels and then rescaling the features in a channel-wise manner. The SE \cite{hu2018squeeze} module selects the most useful feature among channels and improves the effectiveness of the feature representations. In addition to investigating the interdependencies of the channel-wise features, the  spatial correlations are also important for SR tasks. Hu \emph{et al}.\ \cite{hu2019channel} combines channel attention and spatial attention by setting $1\times 1$ trainable convolution filters on each spatial location individually \cite{chen2017sca} together to improve the accuracy. Although incremental results have been obtained, the above methods still have drawbacks. For example, RCAN only focuses on channel-wise attention while ignores the importance of spatial information. Although \cite{hu2019channel} uses the channel and spatial attention mechanisms, their spatial attention mechanism yields high calculation costs. Besides, learning of features in flat image area may require a large image receptive field, while the textual details may require a small one. However, in the above methods, the diversity of features at different scales is ignored. Therefore, it is particularly important to integrate multi-scale information for an attention-based image SR task \cite{shi2017single, hu2018single, li2018multi}.

In this paper, we make a thorough investigation on attention-based image SR methods and then shed light on how simple and effective modifications further improve the state of the art performance. We further propose a unified approach called ``multi-grained attention networks (MGAN)'' to fully exploit the advantages of attention mechanisms in SR tasks, where the ``channel attention'' \cite{zhang2018image, cheng2018sesr, hu2019channel} and ``spatial attention'' \cite{hu2019channel} strategies in previous methods can be essentially considered as two special cases of our method. In our method, the importance of each neuron is computed according to its surrounding regions in a multi-grained fashion and then is used to adaptively re-scale the feature responses.

\begin{figure*}
  \centering
  \includegraphics[width=\linewidth]{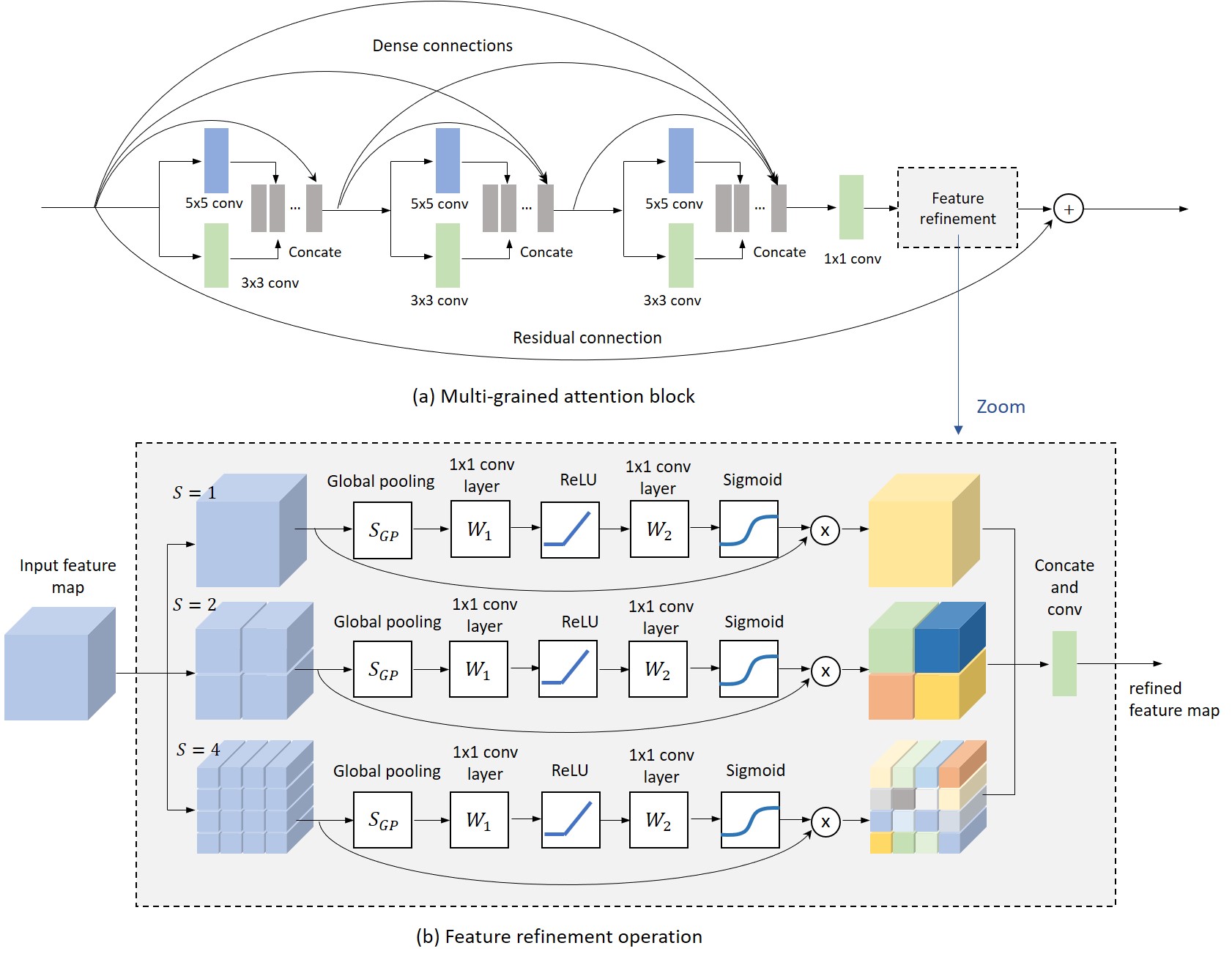}\\
  \caption{(a) The architecture of the basic building block of our method: the Multi-Grained Attention Block (MGAB). (b) The details of the feature refinement operation by using multi-grained attention. $\bigotimes$ denotes element-wise product between two feature maps. The recalibrated features at different scales are further fused into a final representation for further operations.}\label{fig:MGAB}
\end{figure*}

An overview of our method, MGAN, is shown in Fig. \ref{fig:architecture}. In the network, we not only jointly learn feature interdependencies in the channel and spatial dimensions, but also make full use of the skip connections between different layers and scales to improve the information flow during training. The proposed MGAN consists of a series of Multi-Grained Attention Blocks (MGAB). The detail of each block is shown in Fig. \ref{fig:MGAB}, where MGAB produces multi-scale features at each layer and passes to all the subsequent layers through dense connections \cite{huang2017densely}. In each block, the feature responses are adaptively re-scaled in the channel and spatial dimensions to capture richer contextual information, which is beneficial to enhance the discriminability of the network. Specially, we introduce a multi-grained attention structure \cite{wang2018multi} to compute the importance values from the surrounding regions of each neuron in different scales and then individually recalibrates the features at each location (as shown in Fig. \ref{fig:MGAB}(b)). In addition, the output of each MGAB has direct access to a bottleneck layer (with $1\times 1$ convolution) to conduct global feature fusion, which not only boosts the flow of information, but also avoids information loss as the network depth grows.

The contributions of this paper can be summarized as follows.




(1) We introduce the multi-grained attention to a image SR model and show how simple improvement on attention based methods improves the state-of-the-arts. The proposed multi-grained attention mechanism not only captures the importance of feature channels, but also fully exploits spatial context cues.

(2) We further introduce the multi-scale dense connections to fully exploit the features of different layers at multiple scales through dense skip connections. Our model learns richer contextual information and enhances the discriminative representation ability of the features. Ablation studies on benchmark datasets demonstrate the effectiveness of our method. In comparison with some other state-of-the-art SR methods, our method shows superiority in terms of both its accuracy and model size.

The rest of this paper is organized as follows. In Section \ref{sec:Methodology}, we introduce the proposed MGAN in detail. Experimental results and analysis are given in Section \ref{sec:Experiments}. Conclusions are drawn in Section \ref{sec:Conclusion}.

\section{Proposed Method} \label{sec:Methodology}

In this section, we will introduce the proposed multi-grained attention networks in detail.

\subsection{Network Architecture}

As shown in Fig. \ref{fig:architecture}, the processing flow of our method mainly consists of two stages: 1) feature extraction stage, and 2) reconstruction stage. Suppose $I^{LR}$ and $I^{SR}$ represent the input LR image and output SR image, respectively. We firstly apply one convolutional layer to extract the initial features from the input LR image. Then, the initial features are fed into a series of multi-grained attention blocks to produce informative feature maps. Finally, a sub-pixel convolution layer \cite{shi2016real} followed by a convolution layer are used for reconstructing the HR image. The relationship between the $I^{LR}$ image and $I^{SR}$ image can be written as follows:
\begin{equation}
\begin{split}
    I^{SR} &= \mathcal{H}_{MGAN}(I^{LR}, \theta) \\
    &= \mathcal{H}_{up}(\mathcal{H}_{f}(I^{LR})),
\end{split}
\end{equation}
where $\mathcal{H}_{MGAN}$ represents the proposed multi-grained attention  networks, $\mathcal{H}_{f}$ represents the feature extraction operation,  $\mathcal{H}_{up}$ represents the up-sampling and reconstruction operation, and $\theta$ is the parameter of the network to be optimized. To obtain more informative features, we also apply hierarchical feature fusion and global residual learning in the feature extraction pipeline, as shown in Fig. \ref{fig:architecture}. 

Our MGAN is optimized under a standard regression paradigm by minimizing the difference between the reconstructed image $I^{SR}$ and the ground-truth $I^{HR}$. Given a training dataset with $N$ image pairs $\{I^{LR}_i, I^{HR}_i\}_{i=1}^N$, we use the $L_1$ loss function \cite{lim2017enhanced, zhang2018image, hu2019channel, zhang2018residual} to train our model as it introduces less blurring effect. The objective function can be written as
\begin{equation}
    \min_{\theta} L(\theta) = \frac{1}{N}\sum_{i=1}^N \| \mathcal{H}_{MGAN}(I^{LR}_i, \theta) - I^{HR}_i\|_1,
\end{equation}
In the following subsection, we will give more details of the proposed multi-grained attention block.

\subsection{Multi-grained Attention Block}

The architecture of our multi-grained attention block (MGAB) is shown in Fig. \ref{fig:MGAB}. The process flow of each MGAB consists of two stages: 1) feature fusion based on multi-scale dense connections, and 2) feature refinement based on multi-grained attention mechanism. 

To further enhance the network's representation ability and the information flow, the local residual connection is introduced in each attention block. Suppose $F_{d-1}$ and $F_d$ represent the input and output feature representation of the $d$-th MGAB, their relationship can be written by
\begin{equation}
    F_d = F_{d-1} + \mathcal{F}_{att}(F_{d-1}),
\end{equation}
where $\mathcal{F}_{att}$ represents the forward mapping function of a MGAB. In the above operation, we adjust the dimension of the feature maps to make them equal-sized by passing through a set of $1\times 1$ convolutional filters.

\subsubsection{Multi-scale dense connections}

Exploiting multi-scale semantic information of an image is important for effective image representation \cite{zhang2012multi}. The GoogleNet \cite{szegedy2015going} inception module adopts parallel convolutions with different filter sizes to learn multi-scale image representation, leading to state-of-the-art results of the object recognition. Inspired by the GoogleNet \cite{szegedy2015going}, we introduce multi-scale dense connections to our model, which not only learns richer semantic information but also boosts the gradient flow during the training process.

As illustrated in Fig. \ref{fig:MGAB}(a), we introduce two parallel paths in each processing unit with different filter sizes (e.g., $3 \times 3$ and $5 \times 5$). Their outputs of each unit are concatenated together and are then fed to all the subsequent layers in a densely connected fashion, which was introduced by the DenseNet \cite{huang2017densely}. Before the feature refinement stage, we perform local feature fusion among different layers to further extract informative information and make the dimensions of the input and output features the same.





\subsubsection{Squeeze and excitation operation}


Most of the previous methods apply standard convolution to build image SR networks. As the standard convolution operation simply focuses on local regions and cannot obtain long-dependencies over different spatial locations, some recent methods take advantage of the attention mechanism and improve the feature representation of SR methods \cite{zhang2018image, hu2019channel}. 

To introduce the attention mechanism, most previous methods compute the channel-wise global statistics by using average pooling \cite{zhang2018image}, then, the squeeze and excitation (SE) network \cite{hu2018squeeze} performs feature re-calibration by computing the importance of each feature channel. More specifically, let $X=[x_1, x_2, \cdots, x_c]\in \mathbb{R}^{H\times W\times C}$ represents the input features with $C$ feature maps and $H\times W $ size. The SE network first conducts the average pooling operation over global spatial locations of a feature map:
\begin{equation}
    z_k = \mathcal{S}_{GP}(x_k) = \frac{1}{H\times W}\sum_{i=1}^H\sum_{j=1}^W x_k(i,j),
\end{equation}
where $\mathcal{S}_{GP}(\cdot)$ is the global pooling function and $x_k(i,j)$ is the pixel value at position $(i,j)$ of the $k$-th feature channel $x_k$. Then, a gating function is introduced to learn channel-wise interdependencies, which can be written as follows: 
\begin{equation}
    \alpha = \sigma(W_2 \delta(W_1 z)),
\end{equation}
where $\sigma(\cdot)$ and $\delta(\cdot)$ denote the sigmoid and ReLU \cite{glorot2011deep} function, respectively. $W_1$ and $W_2$ are the weights of two $1\times 1$ convolutional layers, which aim to adjust the number of channels and learn the channel importance. $\alpha\in \mathbb{R}^{C\times 1\times 1}$ is the learned attention weights.

Finally, the SE \cite{hu2018squeeze} re-scales the channel feature representations by re-weighting each feature map $x_k$ with the channel-wise attention weights, 
\begin{equation}
    \widetilde x_k = \alpha_k x_k.
\end{equation}

\subsubsection{Multi-grained attention mechanism}

Although the SE network \cite{zhang2018image, hu2018squeeze} learn channel-wise interdependencies effectively, they ignore the diversity of spatial positions, which is crucial for the representation of the image. This is because these methods only exploit all pixels equally to recalibrate channel features without considering the location a priori. For this purpose, we introduce the multi-grained attention mechanism to individually divide the feature maps into different spatial regions and allow different locations to have different weights. 


Specifically, we evenly split the feature map into a set of $S\times S$  spatial regions (e.g. $S=$ 1, 2 or 4), as illustrated in Fig. \ref{fig:MGAB}(b). Each region is then individually processed by passing through the subsequent SE module. Finally, the recalibrated features of each region at different grains are concatenated together to further enhance the feature discriminative presentation. When $S$ is set to 1, the above operation will be reduced to a standard SE module \cite{hu2018squeeze}. In contrast, when we set $S>1$, each spatial region can be learned with a corresponding attention weight. Specially, when each pixel location on the feature map is considered as a unique region, the above operation reduces to the spatial attention as mentioned in \cite{hu2019channel}. 

The Channel-wise and Spatial Attention Residual networks (CSAR) \cite{hu2019channel} is a recent proposed method that is similar to ours, in which they learn individual spatial weights for each pixel location. However, their method suffers from a huge amount of training parameters as they need to learn individual weights for every pixel location in a feature map. As a comparison, our method requires fewer parameters and has lower computational costs. In addition, our method can be considered as a unified approach to introducing attention mechanism at both spatial and channel dimensions, where most methods in previous literature can be viewed as a special case of ours. With the above designs, our method captures rich contextual information in the spatial and channel dimensions at the same time so as to improve the discriminative ability of the model.

\section{Experimental Results and Analysis} \label{sec:Experiments}

\subsection{Datasets and Evaluation Metrics}

In our experiment, we train our model using the DIV2K dataset \cite{timofte2017ntire}, which contains 800 high-quality images. Our model is then tested on five standard benchmark datasets, including Set5 \cite{bevilacqua2012low}, Set14 \cite{zeyde2010single}, Bsd100 \cite{arbelaez2010contour}, Urban100 \cite{huang2015single} and Manga109 \cite{matsui2017sketch} with 4 upscaling factors and two configurations: 1) Bicubic (BI) degradation and 2) Blur-downscale (BD) degradation \cite{zhang2018image, zhang2018residual, zhang2018learning}. The Peak Signal-to-Noise Ratio (PSNR) and the Structural Similarity Index Measurement (SSIM) \cite{wang2004image} are used to evaluate SR performance on Y channel (i.e., luminance) of the transformed YCbCr space.

\begin{table*}
\newcommand{\tabincell}[2]{\begin{tabular}{@{}#1@{}}#2\end{tabular}}  
\centering
\caption{An analysis of the importance of each technical component in our method. The columns $P_1$ - $P_6$ correspond to different experiment configurations. We report the average PSNR on Set5 \cite{bevilacqua2012low} ( $\times {\rm{4}}$ ).}\label{ablation}
\begin{tabular}{l|C{1.5cm}|C{1.2cm}|C{1.2cm}|C{1.2cm}|C{1.2cm}|C{1.2cm}|C{1.2cm}}
\toprule
             & MSRN \cite{li2018multi} & ${P_{\rm{1}}}$ & ${P_{\rm{2}}}$ & ${P_{\rm{3}}}$ & ${P_{\rm{4}}}$ & ${P_{\rm{5}}}$ & ${P_{\rm{6}}}$\\
    \midrule
    Multi-scale dense connections & & \checkmark & \checkmark  & \checkmark  & \checkmark  & \checkmark & \checkmark \\
    \hline
    Hierarchical feature fusion & \checkmark & & \checkmark & \checkmark  & \checkmark  & \checkmark  & \checkmark \\
    \hline
    Channel attention (i.e. S=1) & & & & \checkmark & \checkmark & \checkmark & \checkmark \\
    \hline
    Spatial attention \cite{hu2019channel} & & & & & \checkmark & & \\
    \hline
    Multi-grained attention (S=2) & & & & & & \checkmark & \checkmark   \\
    \hline
    Multi-grained attention (S=4) & & & & & & & \checkmark   \\
    \hline
    PSNR (dB) & 32.25 & 32.23 & 32.31 & 32.37 & 32.39 & 32.42 & 32.45  \\
\bottomrule
 \end{tabular}
\end{table*}
  
\subsection{Implementation Details}

We implement our method using PyTorch \cite{paszke2017automatic} with an NVIDIA Titan Xp GPU. We use data augmentation to increase the diversity of the training images, including random rotation (${90^ \circ }$, ${\rm{18}}{{\rm{0}}^ \circ}$, ${\rm{27}}{{\rm{0}}^ \circ }$) and horizontally flipping. In each training batch, we randomly select 16 LR color patches with the size of $48 \times 48$ as inputs. Our model is trained with Adam \cite{kingma2014adam} optimizer with the parameters ${\beta _{\rm{1}}}{\rm{ = 0}}{\rm{.9}}$, ${\beta _{\rm{2}}}{\rm{ = 0}}{\rm{.999}}$, and $\varepsilon {\rm{ = 1}}{{\rm{0}}^{ - 8}}$. The learning rate is initialized to ${10^{ - 4}}$ and then reduces to half every 200 epochs. In our proposed network, all convolution layers are set with 64 filters and ${\rm{3}} \times {\rm{3}}$ filter size, except for the $5 \times 5$  convolutional layer in MGABs and $1 \times 1$ layer at feature fusion. To keep the size of feature map unchanged during the convolution process, we use zero-padding at the edge of each feature map. In each MGAB, the layer number of ${\rm{3}} \times {\rm{3}}$ and $5 \times 5$ convolutional filters is set the same, which is 3. The reduction ratio in the multi-grained attention unit is set to 16. For upscaling layers, we follow the literatures \cite{lim2017enhanced, zhang2018image, zhang2018residual, shi2016real} and apply ESPCNN \cite{shi2016real} to perform the upsampling operation. The final convolution layer with 3 ${\rm{3}} \times {\rm{3}}$  filters is used to reconstruct the HR output.

\begin{figure*}
  \centering
  \includegraphics[width=\linewidth]{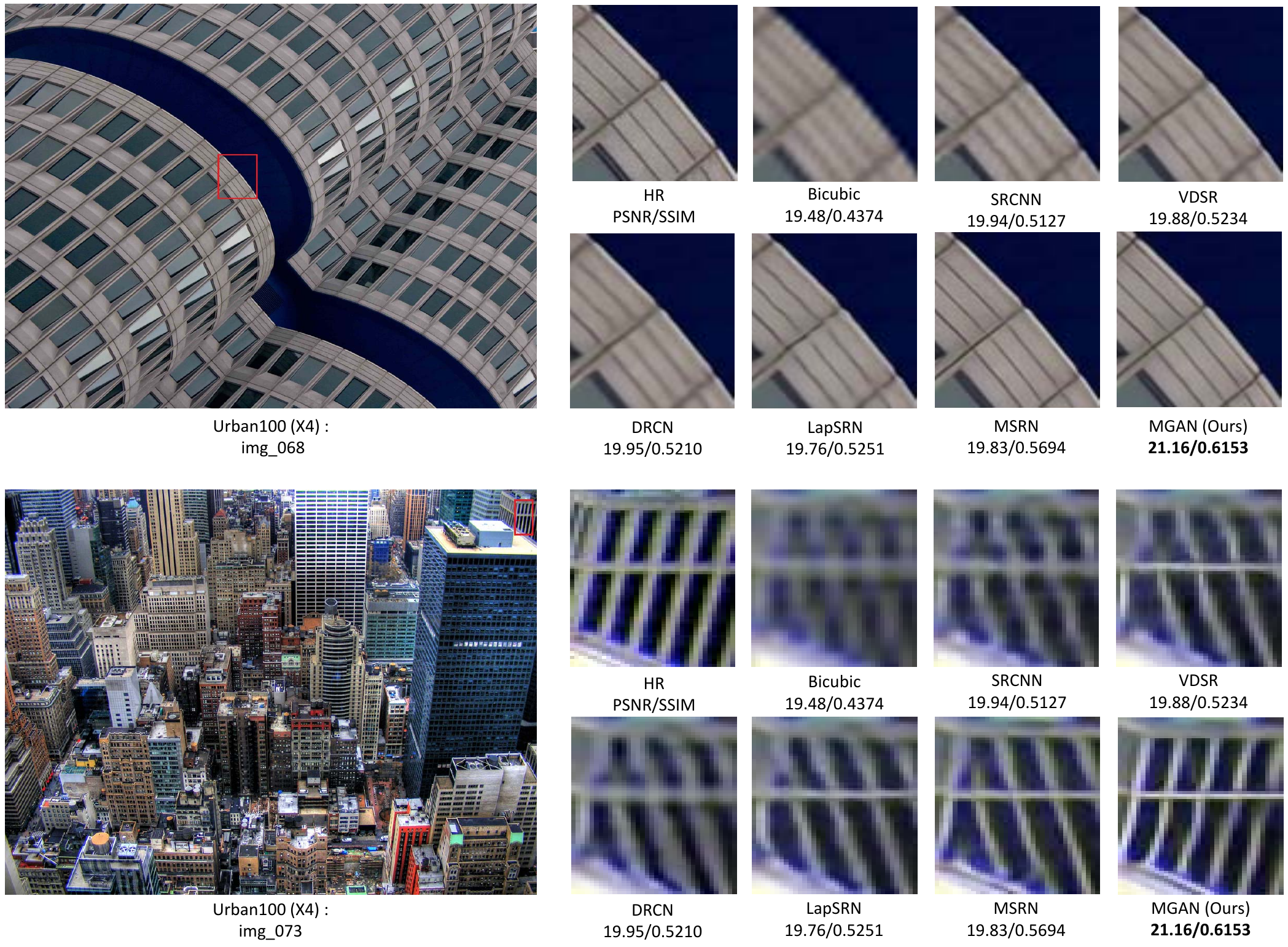}\\
  \caption{A comparison of different SR methods on Urban100 dataset \cite{huang2015single} with $ \times {\rm{4}}$ bicubic degradation. The best result is marked as bold.
}\label{fig:img068073-Urban100}
\end{figure*}

\subsection{Ablation Analysis}

To further evaluate the effectiveness of the proposed methods, the ablation studies are conducted to analyze the importance of each component of the proposed method, including the ``multi-scale dense connections'', ``hierarchical feature fusion'', ``channel attention'', ``spatial attention'', and the ``multi-grained attention''. For fair comparison, we set our MGAN and its variants with the same number of MGAB (8) and convolution filters (64). All evaluations of the ablation analyses are performed based on the same set of configurations. The baseline method ($P_1$ from Table \ref{ablation}) is first evaluated, and then other technical components are gradually added. 

\begin{itemize}
\item \textbf{Multi-scale dense connections}: This corresponds to the multi-scale feature fusion and dense connection in each of our multi-grained attention block. Please refer to Fig.~\ref{fig:MGAB} (a) for more details.
\item \textbf{Hierarchical feature fusion}: This corresponds to the implementations in the method MSRN \cite{li2018multi}.
\item \textbf{Channel attention}: This corresponds to the ``channel attention'' in the method CSAR \cite{hu2019channel}. It can be also considered as a special case of our method when the grid size $S$ is set to 1. Please refer to Fig.~\ref{fig:MGAB} (b) for more details.
\item \textbf{Spatial attention}: This corresponds to the ``spatial attention'' in the method CSAR \cite{hu2019channel}. It can be also considered as a special case of our method when each grid is set to every pixel in the feature map. Please refer to Fig.~\ref{fig:MGAB} (b) for more details.
\item \textbf{Multi-grained attention ($S=2, \ 4$)}: This corresponds to the proposed method with different grid sizes.
\end{itemize}

Table \ref{ablation} shows the ablation study results.
From Table \ref{ablation}, compared with MSRN \cite{li2018multi}, $P_2$ that adds the multi-scale dense connections obtains better SR results, which validates the effectiveness of the multi-scale dense connections. Meanwhile, we can see that the baseline $P_1$ (without hierarchical feature fusion and multi-grained
attention) obtains a relatively low accuracy, where the PSNR only reaches 32.23dB on Set5 $\left( { \times 4} \right)$. Then, we integrate the hierarchical feature fusion and multi-grained attention into the base block, respectively, where we observe a consistent increment of the accuracy. This experiment demonstrates the effectiveness of the proposed hierarchical feature fusion and multi-grained attention. It can also be noticed that the network with channel attention ($P_3$) performs better than those without channel attention ($P_1$ and $P_2$), where the PSNR reaches 32.37 dB. In addition, when we compare the multi-grained attention mechanism ($P_5$ and $P_6$) with other channel and spatial attention mechanisms (i.e., $P_3$ to $P_4$), we can see that our model always achieves better performance than the network with other attention mechanisms. These comparisons consistently confirm the superiority of our method. We also notice that adding more scales to our attention module may not guarantee better performance, but it will greatly increase the computational overhead. To balance the speed and accuracy, we finally set $S = 1,\;2$  and $4$ in our method.


\subsection{Results with Bicubic Degradation (BI)}


We compare our method with several state-of-the-art SR methods, including Bicubic, SRCNN \cite{dong2015image}, VDSR \cite{kim2016accurate}, DRCN \cite{kim2016deeply}, LapSRN \cite{lai2017deep}, EDSR \cite{lim2017enhanced}, and MSRN \cite{li2018multi}\footnote{MSRN has two different implementations and we use its latest results published on its GitHub homepage.}. As suggested by \cite{lim2017enhanced, zhang2018image, zhang2018residual}, we also adopt self-ensemble strategy to further improve our MGAN and accordingly we denote the self-ensembled MGAN as MGAN+. Table \ref{comp_with_others_BI} shows quantitative evaluations with various upscaling factors $\left( { \times 2, \times 3, \times 4, \times 8} \right)$.

Compared to these methods, our MGAN+ achieves higher accuracy on most datasets with various scale factors. Without self-ensemble, our MGAN still obtains favorable results and outperforms most of the other state-of-the-art methods in terms of both PSNR and SSIM. When the up-scaling factor is set to $ \times 4$, the average PSNR gain of our MGAN over the MSRN \cite{li2018multi} are 0.20 dB, 0.11 dB, 0.07 dB, 0.25 dB and 0.24 dB on Set5, Set14, Bsd100, Urban100 and Manga109 datasets, respectively.
Our method achieves better performance on Urban100 and Manga109 datasets mainly because they contain richer structures and details, where the multi-grained attention mechanism in our method shows advantages under such conditions. 

\begin{figure*}
  \centering
  \includegraphics[width=\linewidth]{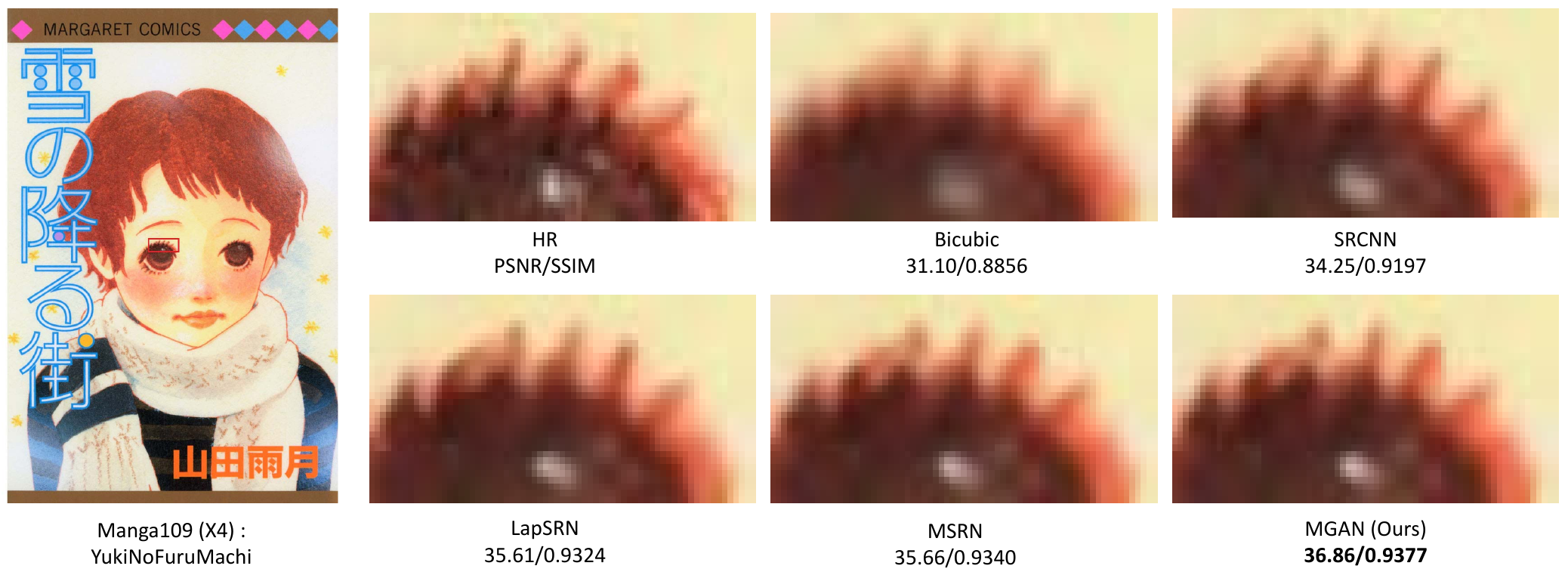}\\
  \caption{
  A comparison of different SR methods on image ``YukiNoFuruMachi'' from the Manga109 dataset \cite{matsui2017sketch} with $ \times {\rm{4}}$ bicubic degradation. The best result is marked as bold.
 }\label{fig:img-Manga109}
\end{figure*}


\begin{figure}
  \centering
  \includegraphics[width=\linewidth]{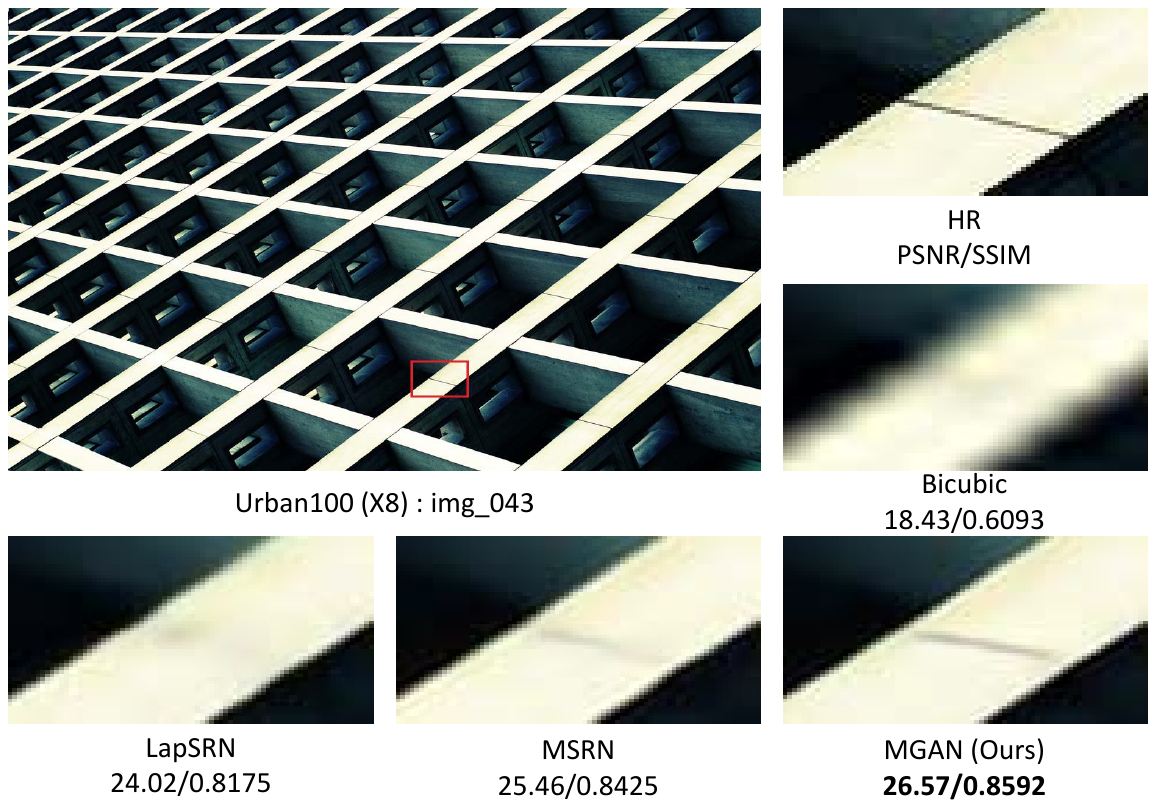}\\
  \caption{
  A comparison of different SR methods on ``img\_043'' from Urban100 dataset \cite{huang2015single} with $ \times {\rm{8}}$ bicubic degradation. The best result is marked as bold.}\label{fig:img043-Urban100}
\end{figure}

To further demonstrate the effectiveness of our method, we provide a visual assessment compared to the other state-of-the-art methods, as shown in Fig.~\ref{fig:img068073-Urban100}, Fig.~\ref{fig:img-Manga109}, and Fig.~\ref{fig:img043-Urban100}. 

In Fig. \ref{fig:img068073-Urban100}, we show visual performance of different methods with a $ \times {\rm{4}}$ upscaling factor  on Urban100 dataset. Our method achieves the best visual results among all methods in terms of both structures and high-frequency details. Using "$image\_073$" as an example, we can see that the Bicubic interpolation generates the worst results with seriously blurring effects along the edges and visually displeasing blurred detailed textures. MSRN \cite{li2018multi} and the other methods generate good results in synthesizing details, but they fail in recovering straight edges and obtain distorted results. As a comparison, our method alleviates the effects of bending along the edges and reconstruct the high-frequency components of the HR images, which is more faithful to the original HR ones. This is because the multi-grained attention mechanism in our method learns not only high-frequency details but also context cues, which is helpful to recover regular shapes and structures in an image. The above results verify the superiority of our MGAN especially for those images with fine structures and details.

 

\subsection{Results with Blur-downscale Degradation (BD)}

As suggested by \cite{zhang2018image, zhang2018residual, zhang2018learning}, we also conduct our experiments with blur-degradation (BD) inputs. We compare our MGAN with 8 state-of-the-art methods: SPMSR \cite{peleg2014statistical}, SRCNN \cite{dong2015image}, FSRCNN \cite{dong2016accelerating}, VDSR \cite{kim2016accurate}, IRCNN \cite{zhang2017learning}, SRMD \cite{zhang2018learning}, RDN \cite{zhang2018residual}, MSRN \cite{li2018multi}. As shown in Table \ref{comp_with_others_BD}, our MGAN outperforms other methods on most datasets with $ \times {\rm{3}}$ upscaling factor. Our enhanced version, MGAN+, further improves the accuracy by using self-ensemble. Particularly, on Urban100 and Manga109, which are more challenging than other datasets, the PSNR gain of the proposed MGAN over MSRN is up to 0.36 dB and 0.37 dB, respectively.

\begin{figure}
  \centering
  \includegraphics[width=\linewidth]{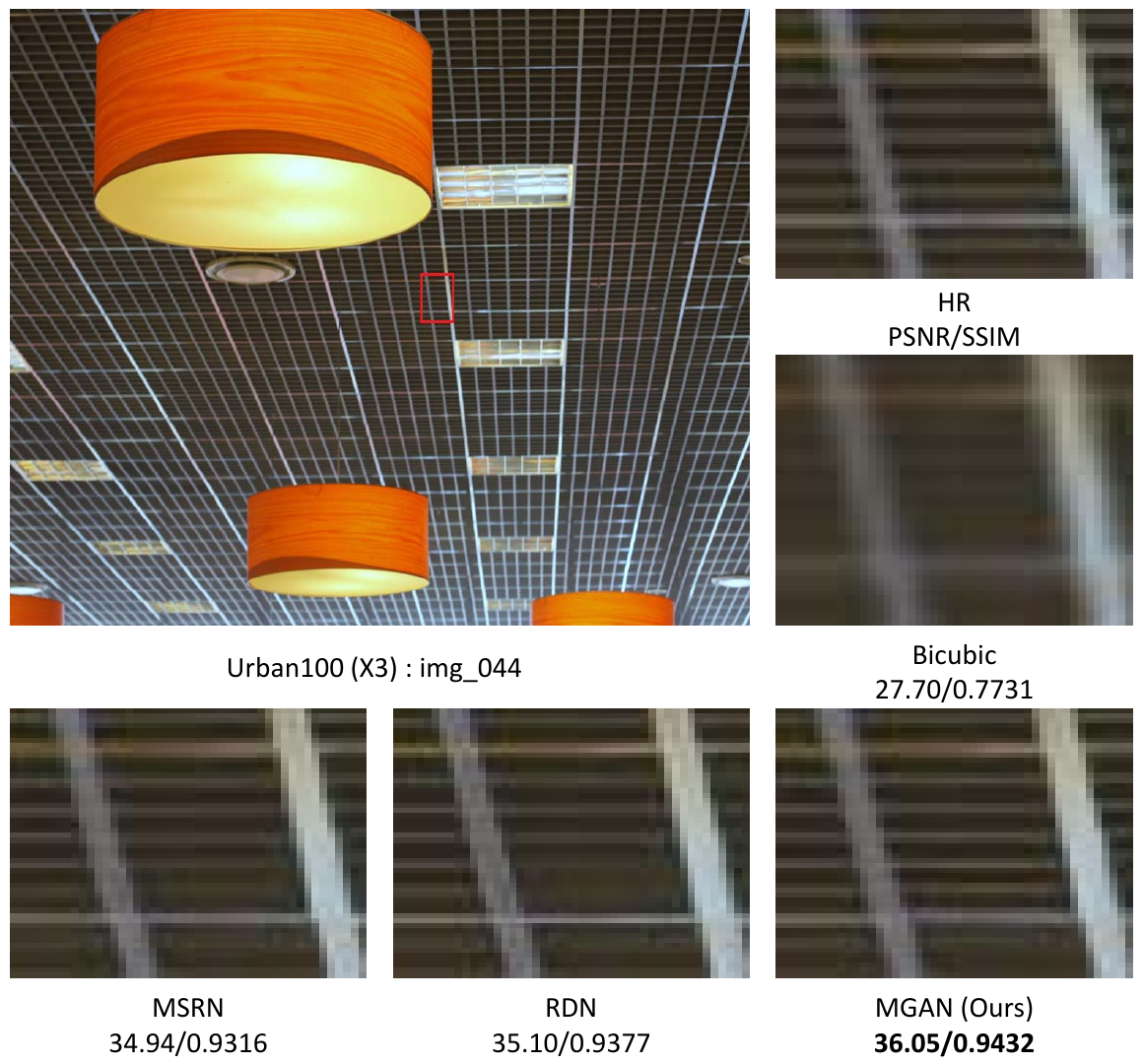}\\
  \caption{A comparison of different SR methods on ``img\_044'' from Urban100 dataset \cite{huang2015single} with $ \times {\rm{3}}$ blur-downscale degradation. The best result is marked as bold.
}\label{fig:img044-Urban100}
\end{figure}

In Fig. \ref{fig:img044-Urban100}, we show visual results on $ \times {\rm{3}}$ up-scaling factor for BD degradation model. For "$image\_044$", RDN and MSRN suffer from the blurring artifacts. As a comparison, our MGAN alleviates the blurring artifacts and recovers more faithful results.


\subsection{Model Size Comparison}

\begin{figure}
  \centering
  \includegraphics[width=\linewidth]{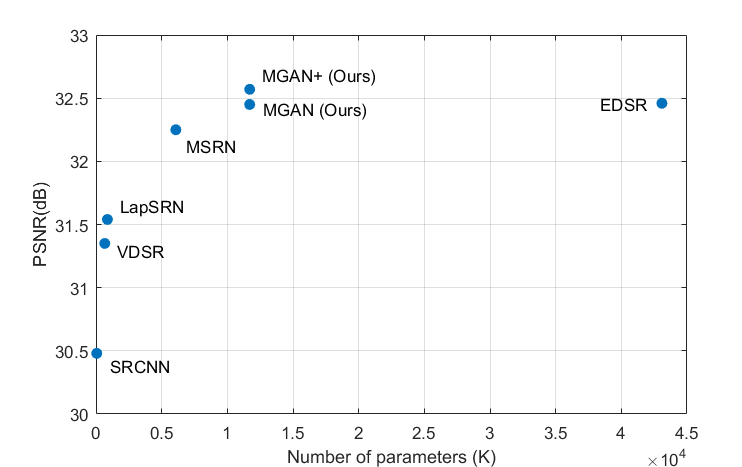}\\
  \caption{Average accuracy (PSNR) of different methods on Set 5 \cite{bevilacqua2012low} with $ \times {\rm{4}}$ bicubic degradation vs. the number of their parameters.}\label{fig:san}
\end{figure}

We further compare the number of parameters and the average PSNR of different methods on Set 5 with upscaling factor $ \times 4$, as shown in Fig. \ref{fig:san}. The results show that our MGAN and MGAN+ achieve a good tradeoff between model size and accuracy. In comparison with other methods, our method (MGAN+) obtains the highest accuracy. Notice that although MGAN achieves comparable performance to EDSR, as shown in Table \ref{comp_with_others_BI}, our MGAN has fewer parameters than EDSR (11.7 M  v.s.  43 M).

\begin{table*}
\newcommand{\tabincell}[2]{\begin{tabular}{@{}#1@{}}#2\end{tabular}}  
\centering
\caption{Quantitative results (PSNR/SSIM) of different SR methods with bicubic degradation. The proposed MGAN and its enhanced version MGAN+ achieve the best SR results for the most datasets and experimental configurations.}\label{comp_with_others_BI}
\begin{tabular}{L{1.8cm}|C{0.8cm}|C{0.8cm}C{0.8cm}|C{0.8cm}C{0.8cm}|C{0.8cm}C{0.8cm}|C{0.8cm}C{0.8cm}|C{0.8cm}C{0.8cm}}
\toprule
\multicolumn{1}{c}{} & \multicolumn{1}{c}{} &
\multicolumn{2}{c}{Set5} & \multicolumn{2}{c}{Set14} & \multicolumn{2}{c}{Bsd100} & \multicolumn{2}{c}{Urban100} & \multicolumn{2}{c}{Manga109}\\
    Methods & Scale & PSNR & SSIM & PSNR & SSIM & PSNR & SSIM & PSNR & SSIM & PSNR & SSIM \\
\midrule
    Bicubic & x2 & 33.66 & 0.9299 & 30.24 & 0.8688 & 29.56 & 0.8431 & 26.88 & 0.8403 & 30.80 & 0.9339 \\
    SRCNN \cite{dong2015image} & x2 & 36.66 & 0.9542 & 32.45 & 0.9067 & 31.36 & 0.8879 & 29.50 & 0.8946 & 35.60 & 0.9663 \\
    VDSR \cite{kim2016accurate} & x2 & 37.53 & 0.9590 & 33.05 & 0.9130 & 31.90 & 0.8960 & 30.77 & 0.9140 & 37.22 & 0.9750 \\
    DRCN \cite{kim2016deeply} & x2 & 37.63 & 0.9588 & 33.04 & 0.9118 & 31.85 & 0.8942 & 30.75 & 0.9133 & 37.63 & 0.9723 \\
    LapSRN \cite{lai2017deep} & x2 & 37.52 & 0.9591 & 33.08 & 0.9130 & 31.08 & 0.8950 & 30.41 & 0.9101 & 37.27 & 0.9740 \\
    EDSR \cite{lim2017enhanced} & x2 & 38.11 & 0.9602 & \textbf{33.92} & 0.9195 & 32.32 & 0.9013 & 32.93 & 0.9351 & 39.10 & 0.9773 \\
    MSRN \cite{li2018multi} & x2 & 38.07 & 0.9608 & 33.68 & 0.9184 & 32.22 & 0.9002 & 32.32 & 0.9304 & 38.64 & 0.9771 \\
    MGAN (ours) & x2 & 38.16 & 0.9612 & 33.83 & 0.9198 & 32.28 & 0.9009 & 32.75 & 0.9340 & 39.11 & 0.9778 \\
    MGAN+ (ours) & x2 & \textbf{38.21} & \textbf{0.9614} & 33.91 & \textbf{0.9205} & \textbf{32.33} & \textbf{0.9015} & \textbf{32.95} & \textbf{0.9354} & \textbf{39.31} & \textbf{0.9782} \\
\midrule
    Bicubic & x3 & 30.39 & 0.8682 & 27.55 & 0.7742 & 27.21 & 0.7385 & 24.46 & 0.7349 & 26.95 & 0.8556 \\
    SRCNN \cite{dong2015image} & x3 & 32.75 & 0.9090 & 29.30 & 0.8215 & 28.41 & 0.7863 & 26.24 & 0.7989 & 30.48 & 0.9117 \\
    VDSR \cite{kim2016accurate} & x3 & 33.67 & 0.9210 & 29.78 & 0.8320 & 28.83 & 0.7990 & 27.14 & 0.8290 & 32.01 & 0.9340 \\
    DRCN \cite{kim2016deeply} & x3 & 33.82 & 0.9226 & 29.76 & 0.8311 & 28.80 & 0.7963 & 27.15 & 0.8276 & 32.31 & 0.9328 \\
    LapSRN \cite{lai2017deep} & x3 & 33.82 & 0.9227 & 29.87 & 0.8320 & 28.82 & 0.7980 & 27.07 & 0.8280 & 32.21 & 0.9350 \\
    EDSR \cite{lim2017enhanced} & x3 & 34.65 & 0.9280 & 30.52 & 0.8462 & 29.25 & 0.8093 & 28.80 & \textbf{0.8653} & 34.17 & 0.9476 \\
    MSRN \cite{li2018multi} & x3 & 34.48 & 0.9276 & 30.40 & 0.8436 & 29.13 & 0.8061 & 28.31 & 0.8560 & 33.56 & 0.9451 \\
    MGAN (ours) & x3 & 34.65 & 0.9292 & 30.51 & 0.8460 & 29.22 & 0.8086 & 28.61 & 0.8621 & 34.00 & 0.9474 \\
    MGAN+ (ours) & x3 & \textbf{34.75} & \textbf{0.9299} & \textbf{30.60} & \textbf{0.8474} & \textbf{29.29} & \textbf{0.8098} & \textbf{28.82} & 0.8651 & \textbf{34.31} & \textbf{0.9490} \\
    \midrule
    Bicubic & x4 & 28.42 & 0.8104 & 26.00 & 0.7027 & 25.96 & 0.6675 & 23.14 & 0.6577 & 24.89 & 0.7866 \\
    SRCNN \cite{dong2015image} & x4 & 30.48 & 0.8628 & 27.50 & 0.7513 & 26.90 & 0.7101 & 24.52 & 0.7221 & 27.58 & 0.8555 \\
    VDSR \cite{kim2016accurate} & x4 & 31.35 & 0.8830 & 28.02 & 0.7680 & 27.29 & 0.7251 & 25.18 & 0.7540 & 28.83 & 0.8870 \\
    DRCN \cite{kim2016deeply} & x4 & 31.53 & 0.8854 & 28.02 & 0.7670 & 27.23 & 0.7233 & 25.14 & 0.7510 & 28.98 & 0.8816 \\
    LapSRN \cite{lai2017deep} & x4 & 31.54 & 0.8850 & 28.19 & 0.7720 & 27.32 & 0.7270 & 25.21 & 0.7560 & 29.09 & 0.8900 \\
    EDSR \cite{lim2017enhanced} & x4 & 32.46 & 0.8968 & 28.80 & \textbf{0.7876} & 27.71 & \textbf{0.7420} & 26.64 & \textbf{0.8033} & 31.02 & 0.9148 \\
    MSRN \cite{li2018multi} & x4 & 32.25 & 0.8958 & 28.63 & 0.7833 & 27.61 & 0.7377 & 26.22 & 0.7905 & 30.57 & 0.9103 \\
    MGAN (ours) & x4 & 32.45 & 0.8980 & 28.74 & 0.7852 & 27.68 & 0.7400 & 26.47 & 0.7981 & 30.81 & 0.9131 \\
    MGAN+ (ours) & x4 & \textbf{32.57} & \textbf{0.8993} & \textbf{28.85} & 0.7874 & \textbf{27.75} & 0.7415 & \textbf{26.68} & 0.8027 & \textbf{31.15} & \textbf{0.9161} \\
    \midrule
    Bicubic & x8 & 24.40 & 0.6580 & 23.10 & 0.5660 & 23.67 & 0.5480 & 20.74 & 0.5160 & 21.47 & 0.6500 \\
    SRCNN \cite{dong2015image} & x8 & 25.33 & 0.6900 & 23.76 & 0.5910 & 24.13 & 0.5660 & 21.29 & 0.5440 & 22.46 & 0.6950 \\
    VDSR \cite{kim2016accurate} & x8 & 25.93 & 0.7240 & 24.26 & 0.6140 & 24.49 & 0.5830 & 21.70 & 0.5710 & 23.16 & 0.7250 \\
    DRCN \cite{kim2016deeply} & x8 & 25.93 & 0.6743 & 24.25 & 0.5510 & 24.49 & 0.5168 & 21.71 & 0.5289 & 23.20 & 0.6686 \\
    LapSRN \cite{lai2017deep} & x8 & 26.15 & 0.7380 & 24.35 & 0.6200 & 24.54 & 0.5860 & 21.81 & 0.5810 & 23.39 & 0.7350 \\
    EDSR \cite{lim2017enhanced} & x8 & 26.96 & 0.7762 & 24.91 & 0.6420 & 24.81 & 0.5985 & 22.51 & 0.6221 & 24.69 & 0.7841 \\
    MSRN \cite{li2018multi} & x8 & 26.95 & 0.7728 & 24.87 & 0.6380 & 24.77 & 0.5954 & 22.35 & 0.6124 & 24.40 & 0.7729 \\
    MGAN (ours) & x8 & 26.90 & 0.7722 & 25.00 & 0.6415 & 24.81 & 0.5979 & 22.47 & 0.6190 & 24.55 & 0.7803 \\
    MGAN+ (ours) & x8 & \textbf{27.09} & \textbf{0.7801} & \textbf{25.16} & \textbf{0.6458} & \textbf{24.91} & \textbf{0.6005} & \textbf{22.69} & \textbf{0.6275} & \textbf{24.87} & \textbf{0.7882} \\
\bottomrule
 \end{tabular}
\end{table*}

\begin{table*}
\newcommand{\tabincell}[2]{\begin{tabular}{@{}#1@{}}#2\end{tabular}}  
\centering
\caption{Quantitative results (PSNR/SSIM) of different SR methods with blur-downscale degradation. The proposed MGAN and its enhanced version MGAN+ achieve the best SR results for all the datasets among other state-of-the-art methods.}\label{comp_with_others_BD}
\begin{tabular}{L{1.8cm}|C{0.8cm}|C{0.8cm}C{0.8cm}|C{0.8cm}C{0.8cm}|C{0.8cm}C{0.8cm}|C{0.8cm}C{0.8cm}|C{0.8cm}C{0.8cm}}
\toprule
\multicolumn{1}{c}{} & \multicolumn{1}{c}{} &
\multicolumn{2}{c}{Set5} & \multicolumn{2}{c}{Set14} & \multicolumn{2}{c}{Bsd100} & \multicolumn{2}{c}{Urban100} & \multicolumn{2}{c}{Manga109}\\
    Methods & Scale & PSNR & SSIM & PSNR & SSIM & PSNR & SSIM & PSNR & SSIM & PSNR & SSIM \\
\midrule
    Bicubic & x3 & 28.78 & 0.8308 & 26.38 & 0.7271 & 26.33 & 0.6918 & 23.52 & 0.6862 & 25.46 & 0.8149 \\
    SPMSR \cite{peleg2014statistical} & x3 & 32.21 & 0.9001 & 28.89 & 0.8105 & 28.13 & 0.7740 & 25.84 & 0.7856 & 29.64 & 0.9003 \\
    SRCNN \cite{dong2015image} & x3 & 32.05 & 0.8944 & 28.80 & 0.8074 & 28.13 & 0.7736 & 25.70 & 0.7770 & 29.47 & 0.8924 \\
    FSRCNN \cite{dong2016accelerating} & x3 & 26.23 & 0.8124 & 24.44 & 0.7106 & 24.86 & 0.6832 & 22.04 & 0.6745 & 23.04 & 0.7927 \\
    VDSR \cite{kim2016accurate} & x3 & 33.25 & 0.9150 & 29.46 & 0.8244 & 28.57 & 0.7893 & 26.61 & 0.8136 & 31.06 & 0.9234 \\
    IRCNN \cite{zhang2017learning} & x3 & 33.38 & 0.9182 & 29.63 & 0.8281 & 28.65 & 0.7922 & 26.77 & 0.8154 & 31.15 & 0.9245 \\
    SRMD \cite{zhang2018learning} & x3 & 34.01 & 0.9242 & 30.11 & 0.8364 & 28.98 & 0.8009 & 27.50 & 0.8370 & 32.97 & 0.9391 \\
    RDN \cite{zhang2018residual} & x3 & 34.58 & 0.9280 & 30.53 & 0.8447 & 29.23 & 0.8079 & 28.46 & 0.8582 & 33.97 & 0.9465 \\
    MSRN \cite{li2018multi} & x3 & 34.50 & 0.9271 & 30.43 & 0.8427 & 29.15 & 0.8060 & 28.15 & 0.8513 & 33.74 & 0.9447 \\
    MGAN (ours) & x3 & 34.63 & 0.9284 & 30.54 & 0.8450 & 29.24 & 0.8081 & 28.51 & 0.8580 & 34.11 & 0.9467 \\
    MGAN+ (ours) & x3 & \textbf{34.73} & \textbf{0.9290} & \textbf{30.64} & \textbf{0.8462} & \textbf{29.30} & \textbf{0.8091} & \textbf{28.70} & \textbf{0.8610} & \textbf{34.42} & \textbf{0.9483} \\
\bottomrule
 \end{tabular}
\end{table*}

\section{Conclusion} \label{sec:Conclusion}

We made a thorough investigation on attention mechanisms in a super-resolution model and show how simple improvements on previous attention-based model improve the accuracy. We hereby propose a method called ``multi-grained attention networks (MGAN)'' which fully exploits the advantages of attention mechanisms in SR tasks. We found that both ``channel attention'' and ``spatial attention'' are essential for a SR model, and the multi-grained attention integration can further boost the accuracy of current state-of-the-art models. Besides, the proposed multi-scale dense connections allows MGAN not only to capture the image features in different scales, but also make full use of different layers of information. Extensive experiments and ablation studies on a variety of super-resolution models demonstrate the superiority of the proposed methods in terms of quantitative results and visual appearance.

\appendices

\ifCLASSOPTIONcaptionsoff
  \newpage
\fi 

\bibliographystyle{IEEEtran}   
\bibliography{egbib}   

\end{document}